\documentclass[11pt, a4paper, copyright]{perplexity}
\usepackage{xspace}
\usepackage{xcolor}
\makeatletter
\newcommand{\ours@name}{pplx-\allowbreak embed-\allowbreak v2-\allowbreak late}
\DeclareRobustCommand{\ours}{\@ifnextchar0{\ours@small}{\@ifnextchar9{\ours@large}{\texttt{\ours@name}}}}
\def\ours@small0_6b{\texttt{\ours@name-\allowbreak 0.6b}}
\def\ours@large9b{\texttt{\ours@name-\allowbreak 9b}}
\makeatother

\usepackage[authoryear, sort&compress, round]{natbib}
\providecommand{\doi}[1]{\href{https://doi.org/#1}{doi:\,\nolinkurl{#1}}}

\usepackage{graphicx}
\usepackage{multirow}                                                                               
\usepackage{float}                                                                                   
\usepackage{adjustbox}
\usepackage{tablefootnote}

\uselogo{} 

\title{pplx-embed-v2-late Technical Report}

\author[*,1]{Markus Krimmel}
\author[*,1]{Louis Milliken}
\author[1]{Ekaterina Glazkova}
\author[1]{Maximilian Schall}
\author[1]{Antoine Chaffin}
\author[1]{Sedigheh Eslami}
\author[1]{Jonas Golde}
\author[1]{Denis Bykov}
\author[1]{Claire Beamer}
\author[1]{Jeremy Yang}
\author[1]{Bo Wang}

\affil[*]{Equal contributions}
\affil{Perplexity AI}

\begin{abstract}
In this report, we introduce \ours, a pair of 0.6B and 9B multilingual and multimodal late-interaction models.
Both models achieve state-of-the-art results within their respective size classes across a range of text, image, and visual document retrieval benchmarks. On visual document retrieval tasks, \ours0_6b is competitive with models several times larger than itself. The models are trained via representation distillation from an 18B teacher. Consequently, they share an embedding space that supports asymmetric deployment: documents can be indexed with the 9B model and queried with the low-latency 0.6B model. 
Both models are available at \url{https://huggingface.co/collections/perplexity-ai/pplx-embed-v2}.

\end{abstract}

\begin{document}

\maketitle

\section{Introduction}
Retrieval systems face a trade-off between efficiency and the depth of interaction between queries and documents. Dense embeddings interact via cheap inner products, which allow for efficient approximate nearest neighbor search. At the other extreme, cross-encoders process the query and document together, enabling richer interaction but requiring a separate forward pass for every ranked candidate. Late-interaction models~\citep{colbertv1} occupy the space between the two. Documents are indexed offline and represented as bags of embedding vectors. These document vectors interact with query vectors via the MaxSim rule, allowing richer interaction than dense embeddings~\citep{provablymoreexpressive}.
Late-interaction models have shown especially strong performance in out-of-domain settings~\citep{colbertv2, jacolbertv2}, long-context retrieval~\citep{modernbert,GTE-ModernColBERT,colbert-edge}, agentic search~\citep{Agent-ModernColBERT, wholembed_v3}, and visual document retrieval (VDR)~\citep{colpali, jina-v4}. 

In this report, we introduce \ours, a pair of 0.6B and 9B multimodal and multilingual late-interaction models. Both models are distilled at the representation level from a contrastively trained 18B teacher. We find that representation distillation outperforms contrastive training alone, especially at smaller model scales.
Since the 0.6B and 9B models are trained to match the same teacher, the two encoders are interchangeable: a query embedded by the 0.6B model can be scored directly against an index built by the 9B model, and vice versa. This allows for hybrid deployment options where the latency-critical query path can be served by the fast 0.6B model while documents are indexed offline via the stronger 9B model, with the cost of indexing amortized over the lifetime of the index.
Both models individually achieve state-of-the-art performance on a range of text, image, and VDR benchmarks, including ViDoRe-V3~\citep{vidorev3}.
In concurrent work, \citet{colnanovdr} study representation-level distillation in late-interaction models to produce small query encoders. Our work distills query and document token representations into compact multimodal late-interaction models that can serve as both query and document encoders.

\section{Training}
In the following, we describe the training procedure for the 18B teacher model and the 9B and 0.6B students.
The teacher model is trained contrastively via an InfoNCE~\citep{oord2019representationlearningcontrastivepredictive} objective in two stages: first on query-document pairs, then on triplets of queries, relevant documents, and hard negatives.
The students are trained via a pointwise distillation objective similar to that of LEAF~\citep{leaf} to match the embeddings of the teacher. We reuse the pair data from the contrastive training stage for distillation. All of our late-interaction models have 128-dimensional embeddings.

\subsection{Contrastive Teacher Training}
We use Qwen3.5~\citep{qwen3-5} as the backbone for our late-interaction models. Qwen3.5 uses a hybrid architecture consisting of gated full attention layers with a causal mask and recurrent Gated DeltaNet~\citep{gated-deltanet} (GDN) layers. We disable the causal mask in the full attention layers, yielding a hybrid bidirectional-causal architecture. We initialize the 18B late-interaction model by pruning the last 20 layers from Qwen3.5-27B. This reduces the cost of distillation while maintaining a strong training signal.

\begin{figure}
    \centering
    \includegraphics[width=\linewidth]{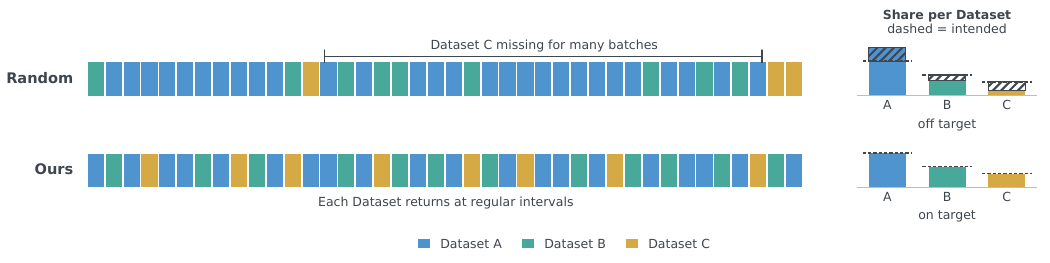}
    \caption{Comparison of our sampling method with independent random sampling according to the dataset weights.}
    \label{fig:kronecker}
\end{figure}

\paragraph{Pair Training.} Our pair data is composed of 28 datasets which we further partition by language and domain to increase the difficulty of in-batch negatives. This yields 981 sub-datasets with 186M rows of query-document pairs in total, 88.3\% of which are text-text pairs, 8.3\% are text-image pairs and 3.4\% are text-visual document pairs. Of the text-text pairs, 14M are synthesized by generating queries for specific passages in PDF documents from the FinePDFs~\citep{finepdf} dataset.
During training, we adjust the sampling weights of each dataset to reach a data mixture of 56.5\% text-text pairs 30.9\% text-image pairs, and 12.6\% text-visual document pairs. Following previous work~\citep{gunther2023jinaembeddingsnovelset}, we sample full batches from the sub-datasets. Since many sub-datasets are domain or language slices of the same source dataset, we introduce a novel quasi-random sampling strategy to ensure that sub-datasets are sampled both~(i)~in proportions close to the intended mixture and (ii)~at regular intervals across training steps, avoiding both long gaps and temporal clustering. The sampling should satisfy these properties not only at the level of sub-datasets but also at the source-dataset level. We achieve this by constructing a cumulative sampling distribution in which all sub-datasets from the same source occupy contiguous intervals and mapping a randomly shifted golden-ratio Kronecker sequence through its inverse CDF.
The sequence’s low-discrepancy coverage~\citep[Chapter~2, Theorem~3.4]{kuipers1974uniform} and structured return times~\citep{slater1950distribution} apply at both levels, guaranteeing proportional, well-spaced sampling of individual slices and their parent sources. We illustrate this in Figure~\ref{fig:kronecker}.

We train the model with LoRA~\citep{lora} on all attention and feed-forward projections to stabilize training and mitigate catastrophic forgetting. The adapters have rank 128. 
Our training objective is the InfoNCE loss with a contrastive batch size of 16,384 query-document pairs. Following~\citep{qwen3-embedding}, we mask in-batch negatives based on a margin of 0.2 to the query-positive MaxSim score. During training, we truncate queries to 256 tokens and documents to 1,000 tokens, and resize images to at most 917,504 pixels while preserving their original aspect ratios.
To enable the large contrastive batch size, we use gradient caching~\citep{gradient-caching} and implement the MaxSim operation as a custom Triton~\citep{triton} kernel to avoid materializing token-level similarity matrices. This work was concurrent with the efficient MaxSim implementations by~\citet{pony2026flashmaxsim,sharma2026tilemaxsim,late_interaction_kernels_2026}.

\paragraph{Triplet Training.} After pair training, we merge the LoRA weights into the backbone and add a new adapter with rank 32. We freeze the vision tower and train only on text triplets. We use at most four hard negatives per query and contrast each positive against its corresponding hard negatives and all other in-batch documents. Our triplet data is composed of public triplet datasets, mined hard negatives, and synthetic triplets created by modifying key information in highly templated documents.

\subsection{Representation Distillation}
To distill our 18B teacher model down to 0.6B and 9B student models, we train the students to reproduce the teacher’s token-level representations by minimizing the Euclidean distance between corresponding $\ell_2$-normalized embeddings. We average these distances over unmasked query and document positions separately, then combine the two averages with equal weight. The teacher remains frozen, so gradients update only the student. We let $\mathcal Q$ and $\mathcal D$ denote the unmasked query-token and document-token positions across the batch and indicate student and teacher embeddings by superscripts $S$ and $T$. With $\mathbf{q}_{it}$ and $\mathbf{d}_{it}$ denoting query and document embeddings at token position $t$ of example $i$ within the batch, we formulate the loss as:
\begin{equation*}
    \mathcal{L} =
    \frac{1}{|\mathcal{Q}|}
    \sum_{(i,t)\in\mathcal{Q}}
    \left\|\mathbf{q}_{it}^{\mathrm{S}}-\mathbf{q}_{it}^{\mathrm{T}}\right\|_2
    +
    \frac{1}{|\mathcal{D}|}
    \sum_{(i,t)\in\mathcal{D}}
    \left\|\mathbf{d}_{it}^{\mathrm{S}}-\mathbf{d}_{it}^{\mathrm{T}}\right\|_2.
\end{equation*}
Since this loss is not contrastive, we can mix sub-datasets when sampling a batch to reduce gradient noise without weakening our training signal.

\paragraph{Distillation to 9B.} We use Qwen3.5-9B with bidirectional full attention as the backbone for our 9B late-interaction model. Unlike during contrastive training of the 18B model, LoRA does not appear to provide sufficient capacity to achieve satisfactory downstream performance with representation distillation. We hypothesize that this is caused by the constraints imposed on the embedding space by the distillation objective.
To increase capacity while limiting catastrophic forgetting, we fully fine-tune the last eight layers of the text tower while applying LoRA to the remaining projections.

\paragraph{Distillation to 0.6B.} The 0.6B late-interaction model is initialized from a pruned version of Qwen3.5-0.8B with bidirectional full attention layers.
Within each block of Qwen3.5-0.8B, we prune two of the three GDN layers, producing an architecture that alternates between gated full attention and GDN layers.
The resulting model has 594M parameters in total, of which 100M are used for image encoding and 254M are in the token-embedding table. Hence, only about 240M parameters are active for text encoding and 340M for image encoding. This small size allows deployment on edge devices.
We find that, especially on VDR tasks, full fine-tuning of the 0.6B backbone outperforms parameter-efficient approaches. Hence, in contrast to the 18B and 9B models, we fine-tune all parameters.

\section{Evaluation}
We evaluate on various text-only and text-to-image benchmarks. We present results on single-hop and agentic retrieval tasks.
We excluded any training data that overlaps with evaluated tasks.

\subsection{Text retrieval}
For text retrieval, we evaluate a collection of 72 public tasks across six domains: finance (18 tasks), legal (20), health (8), conversation (5), tech (4), and multilingual (17), which we bundle into an internal \emph{domain-specific benchmark}. Additionally, we evaluate the text version of ViDoRe-V3~\citep{vidorev3} and the Q2D-Web~\citep{q2d-web} benchmark.
Text baselines include Nemotron 3 Embed 8B~\citep{nemotron3embed8b}, Qwen3 Embedding~\citep{qwen3-embedding}, Gemini Embedding 2~\citep{gemini-embedding-2}, and Voyage 4~\citep{voyage4}.

\paragraph{Domain-Specific Benchmark.} The domain-specific collection is built from public datasets such as FinQA~\citep{finqa}, AILACasedocs~\citep{aila-casedocs} and XLSum~\citep{xlsum}. We do not disclose the full task list to reduce the risk of future models being tuned to it; as a result, these scores cannot be independently reproduced.
\ours9b averages 81.3\% nDCG@10, 1.6~pp above the next best model, and scores highest on every domain except health, where nemotron-3-embed-8b and voyage-4-large score higher. \ours0_6b averages 78.0\% and scores highest among models with fewer than 1B parameters on every domain (Table~\ref{tab:eval-domain}).
\begin{table}[t]
\centering
\caption{Domain-specific text retrieval performance (nDCG@10, \%). The average weights the six domains equally; parentheses show the number of tasks in each domain. Dim.\ is the embedding dimension (per token for multi-vector models). Best score per column in bold.}
\label{tab:eval-domain}
\small
\setlength{\tabcolsep}{4pt}
\resizebox{\linewidth}{!}{%
\begin{tabular}{l|cc||c|cccccc}
\toprule
Model & Dim. & Vectors & Avg. & Finance (18) & Legal (20) & Health (8) & Conv. (5) & Tech (4) & Multiling. (17) \\
\midrule
\multicolumn{10}{l}{\textit{>1B / size unreported}} \\
\texttt{pplx-embed-v2-late-9b} & 128 & multi & \textbf{81.3} & \textbf{82.5} & \textbf{81.6} & 82.5 & \textbf{81.5} & \textbf{70.4} & \textbf{89.5} \\
\texttt{nemotron-3-embed-8b} & 4096 & single & 79.7 & 80.8 & 81.1 & \textbf{83.9} & 77.4 & 67.5 & 87.9 \\
\texttt{gemini-embedding-2} & 3072 & single & 78.3 & 77.0 & 78.5 & 82.4 & 76.2 & 67.7 & 88.1 \\
\texttt{voyage-4-large} & 1024 & single & 79.0 & 81.5 & 80.0 & 83.0 & 76.7 & 65.1 & 87.9 \\
\texttt{qwen3-embedding-8b} & 4096 & single & 74.8 & 72.3 & 76.8 & 80.2 & 72.0 & 62.7 & 84.7 \\
\midrule
\multicolumn{10}{l}{\textit{<1B}} \\
\texttt{pplx-embed-v2-late-0.6b} & 128 & multi & 78.0 & 79.9 & 78.2 & 78.0 & 79.7 & 66.6 & 85.8 \\
\texttt{voyage-4-nano} & 2048 & single & 74.6 & 77.3 & 75.8 & 76.0 & 74.7 & 59.4 & 84.3 \\
\texttt{qwen3-embedding-0.6b} & 1024 & single & 68.4 & 67.9 & 70.0 & 72.4 & 68.4 & 53.9 & 77.8 \\
\bottomrule
\end{tabular}}
\end{table}

\paragraph{Q2D-Web.} The Q2D-Web benchmark contains about 70,000 agent-reformulated queries from production search traffic over a corpus of 190M web documents. Because it targets first-stage retrieval, we report Recall@1000. Relevance is judged in three ways: documents cited by the agent (Citation), documents ranked highly by our production retrieval system (Web), and the union of both extended with LLM judgments (Combined). We focus on Combined, which reduces bias towards models that rank similarly to the production system. On Combined, the 9B and 0.6B models score 74.8\% and 73.6\%, compared with 69.3\% for the strongest baseline (Table~\ref{tab:eval-q2d-web}).

\begin{table}[ht]
\centering
\caption{Text retrieval on the public subset of ViDoRe-V3 after converting page images to Markdown with OCR (average nDCG@10, \%) and on Q2D-Web (Recall@1000, \%) under three relevance judgment sets. Dim.\ is the embedding dimension (per token for multi-vector models). \texttt{topk-embed-v1} ViDoRe-V3 scores are taken from the model cards~\citep{topk-embed-v1-small,topk-embed-v1-xsmall} at 128 dimensions; on Q2D-Web, these models were evaluated with an output dimension of 128. -- denotes not evaluated. Best score per column in bold.}
\label{tab:eval-q2d-web}
\label{tab:eval-vidore-md}
\small
\setlength{\tabcolsep}{3pt}
\begin{tabular}{l||cc||c|ccc}
\toprule
 & & & ViDoRe-V3 (text) & \multicolumn{3}{c}{Q2D-Web} \\
Model & Dim. & Vectors & Avg. & Citation & Web & Combined \\
\midrule
\multicolumn{7}{l}{\textit{>1B / size unreported}} \\
\texttt{pplx-embed-v2-late-9b} & 128 & multi & \textbf{64.7} & \textbf{88.2} & \textbf{68.8} & \textbf{74.8} \\
\texttt{nemotron-3-embed-8b} & 4096 & single & 60.6 & 83.1 & 61.7 & 69.3 \\
\texttt{topk-embed-v1-2b} & 128 & multi & 59.2 & 79.2 & 57.7 & 65.3 \\
\texttt{gemini-embedding-2} & 3072 & single & 58.9 & -- & -- & -- \\
\texttt{voyage-4-large} & 1024 & single & 58.9 & -- & -- & -- \\
\texttt{qwen3-embedding-8b} & 4096 & single & 51.5 & 79.3 & 58.3 & 65.3 \\
\midrule
\multicolumn{7}{l}{\textit{<1B}} \\
\texttt{pplx-embed-v2-late-0.6b} & 128 & multi & 61.2 & 86.7 & 67.3 & 73.6 \\
\texttt{topk-embed-v1-0.8b} & 128 & multi & 56.9 & 77.4 & 56.2 & 63.6 \\
\texttt{voyage-4-nano} & 2048 & single & 53.3 & 69.5 & 50.1 & 57.4 \\
\texttt{qwen3-embedding-0.6b} & 1024 & single & 44.2 & 71.2 & 51.8 & 58.7 \\
\bottomrule
\end{tabular}
\end{table}

\paragraph{ViDoRe-V3.} On the Markdown version of ViDoRe-V3 (Table~\ref{tab:eval-vidore-md}), both \ours{} models outperform all evaluated baselines, with the 9B model more than 4~pp above the best competitor.

\subsection{Text-to-image retrieval}
\paragraph{ViDoRe-V3.} Our primary VDR benchmark is the image version of ViDoRe-V3. We compare against dense (qwen3-vl-embed-8b~\citep{qwen3-vl-embedding}, gemini-embedding-2) and late-interaction (topk-embed-v1 0.8B and 2B) vision-language models, and against two vision-only late-interaction model families trained specifically for image retrieval (EVIE~\citep{evie} and Nemotron ColEmbed V2~\citep{nemotron-colembed-v2}).
Among vision-language models, \ours9b and \ours0_6b have the two highest averages. \ours9b scores 65.2\%, below EVIE-8B (66.4\%) and EVIE-4.5B (65.9\%), which have output dimensions of 4,096 and 2,048 per token, respectively, compared to \ours's output dimension of 128. \ours0_6b scores 62.3\%, 0.9~pp below topk-embed-v1-2b at its full width of 2,048 dimensions per token and 0.4~pp above it at 128 dimensions, 3.3~pp higher than topk-embed-v1-0.8b at 128 dimensions.

\begin{table}[ht]
\centering
\caption{Text-to-image retrieval (nDCG@10, \%) on the public subset of ViDoRe-V3 (average), MIRACL-Vision and PPLX-Q2I. Vision-only models are trained specifically for image retrieval and are listed separately from vision-language models. Dim.\ is the embedding dimension (per token for multi-vector models). Nemotron ColEmbed and EVIE models were not evaluated on PPLX-Q2I because their high output dimensionality makes indexing impractical; -- denotes not evaluated. Best score per column in bold.}
\label{tab:eval-vidore-img}
\label{tab:eval-additional-image}
\small
\setlength{\tabcolsep}{5pt}
\resizebox{\linewidth}{!}{%
\begin{tabular}{l||cc||c|c|c}
\toprule
Model & Dim. & Vectors & ViDoRe-V3 (image) & MIRACL-Vision & PPLX-Q2I \\
\midrule
\multicolumn{6}{l}{\textit{Vision + language, >1B / size unreported}} \\
\texttt{pplx-embed-v2-late-9b} & 128 & multi & 65.2 & 75.7 & 65.0 \\
\texttt{topk-embed-v1-2b} & 128 & multi & 61.9 & 60.2 & 51.5 \\
\texttt{qwen3-vl-embedding-8b} & 4096 & single & 58.3 & 68.5 & 50.6 \\
\texttt{gemini-embedding-2} & 3072 & single & 46.0 & \textbf{77.5} & \textbf{67.0} \\
\midrule
\multicolumn{6}{l}{\textit{Vision + language, <1B}} \\
\texttt{pplx-embed-v2-late-0.6b} & 128 & multi & 62.3 & 67.5 & 60.4 \\
\texttt{topk-embed-v1-0.8b} & 128 & multi & 59.0 & 56.5 & 42.3 \\
\midrule
\multicolumn{6}{l}{\textit{Vision-only}} \\
\texttt{nemotron-colembed-v2-8b} & 4096 & multi & 63.5 & 68.6 & -- \\
\texttt{nemotron-colembed-v2-4b} & 2560 & multi & 61.4 & 62.7 & -- \\
\texttt{evie-8b} & 4096 & multi & \textbf{66.4} & -- & -- \\
\texttt{evie-4.5b} & 2048 & multi & 65.9 & -- & -- \\
\bottomrule
\end{tabular}}
\end{table}

\paragraph{MIRACL-Vision.} The MIRACL-Vision benchmark~\citep{miracl-vision} consists of a corpus of Wikipedia article screenshots. On this task, the vision-only Nemotron ColEmbed V2 4B and 8B models score 62.7\% and 68.6\%, respectively; \ours0_6b scores 67.5\% and \ours9b scores 75.7\%, 1.8 percentage points below the highest-scoring model, gemini-embedding-2 (Table~\ref{tab:eval-additional-image}).

\paragraph{PPLX-Q2I.} PPLX-Q2I is built from production image-search logs, with 10,000 queries and 100,000 images (both visual documents and natural images); a query's relevant images are those served for it in production. The 0.6B and 9B \ours~models score 60.4\% and 65.0\%, compared with 67.0\% for gemini-embedding-2 and 50.6\% for qwen3-vl-embedding-8b (Table~\ref{tab:eval-additional-image}).

\begin{figure}[!htb]
    \centering
    \includegraphics[width=\linewidth]{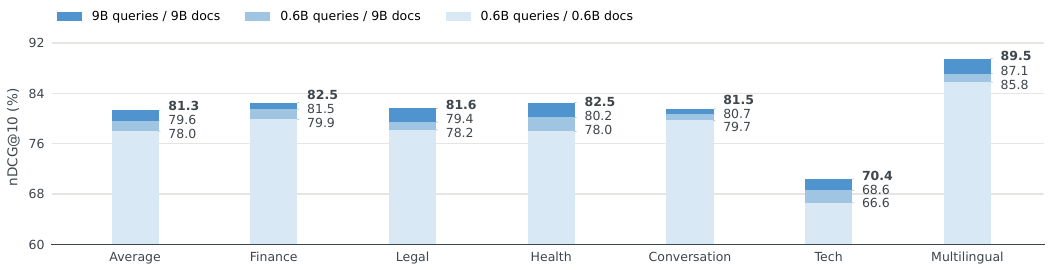}
    \caption{Cross-model retrieval nDCG@10 for three query/document model pairings. Results across six text domains.}
    \label{fig:eval-cross-model}
\end{figure}

\subsection{Cross-model retrieval}
Because the two models share an embedding space, a corpus can be indexed with the 9B model and queried with the 0.6B model, keeping query encoding cost the same as in the 0.6B$\rightarrow$0.6B setting. We illustrate this in Figure~\ref{fig:eval-cross-model} on the domain-specific benchmarks. The 0.6B$\rightarrow$9B setting scores higher than 0.6B$\rightarrow$0.6B on every domain, by 1.6~pp on the domain average. The 9B$\rightarrow$9B setting scores highest at 81.3\%. The hybrid setting thus improves retrieval performance without increasing query encoding cost.

\subsection{Agentic retrieval}
\begin{figure}[!htb]
    \centering
    \includegraphics[width=\linewidth]{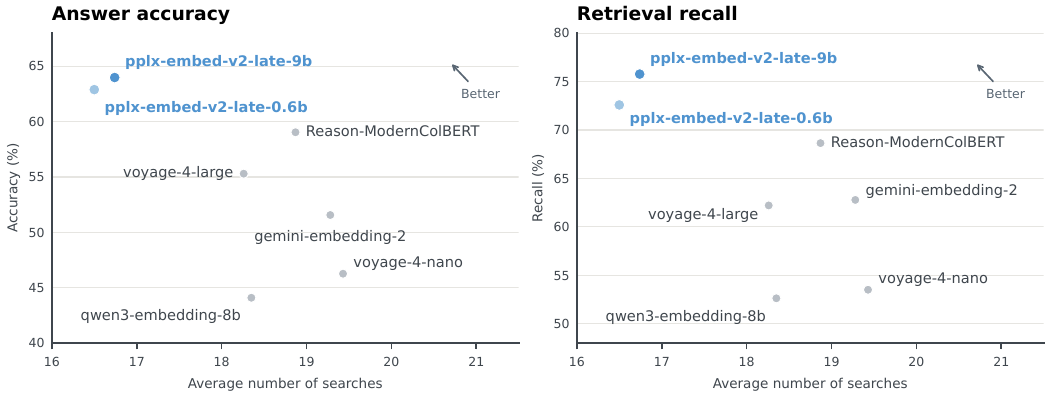}
    \caption{BrowseComp+ accuracy and recall versus the average number of searches (GPT-OSS-120B agent, high effort).}
    \label{fig:eval-browsecomp}
\end{figure}

In BrowseComp+~\citep{browsecompplus}, an LLM agent (GPT-OSS-120B~\citep{gpt-oss}, high effort) answers complex queries by searching a corpus with the embedding model under test, and a second LLM judges the answers. Both \ours{} models achieve higher accuracy with fewer searches than all evaluated baselines (Figure~\ref{fig:eval-browsecomp}). The 9B model reaches 64.0\% accuracy, 4.9~pp above Reason-ModernColBERT~\citep{Reason-ModernColBERT}, the only other late-interaction model evaluated, and 8.7~pp above the best dense model, voyage-4-large.

\FloatBarrier

\section{Conclusion}
We presented \ours{}, a pair of 0.6B and 9B multimodal, multilingual late-interaction models distilled at the representation level from a contrastively trained 18B teacher.
On our domain-specific text benchmarks, Q2D-Web and the Markdown version of ViDoRe-V3, both models outperform all evaluated baselines compared, to other models of their size. On ViDoRe-V3 image retrieval, they have the highest averages among vision-language models, while vision-only EVIE models score higher.
Because both students are trained to match the same teacher, they share an embedding space: indexing with the 9B model and querying with the 0.6B model improves over using the 0.6B model alone, at the same query encoding cost.
In agentic retrieval, both models reach higher BrowseComp+ accuracy with fewer searches than all evaluated baselines.

\bibliography{main}

\end{document}